\documentclass[]{spie}  %>>> use for US letter paper
\pdfoutput=1
\usepackage[]{graphicx}

\usepackage{upgreek}
\usepackage{color}
\usepackage{hyperref}

\newcommand{\aspconf}{ASP Conf.\ Ser.}
\newcommand{\mnras}{MNRAS}
\newcommand{\pasp}{PASP}
\newcommand{\apj}{ApJ}
\newcommand{\apjs}{ApJS}
\newcommand{\aap}{A\&A}

\newcommand{\ascl}[1]{\href{http://www.ascl.net/#1}{ascl:#1}}

\newcommand{\micron}{\,$\upmu$m}
\newcommand{\secref}[1]{\S\ref{#1}}

\title{An overview of the planned CCAT software system}

\author{Tim~Jenness\supit{a},
Martin~C.~Shepherd\supit{b},
Reinhold~Schaaf\supit{c},
Jack~Sayers\supit{b},
Volker~Ossenkopf\supit{d},
Thomas~Nikola\supit{a},
Gaelen~Marsden\supit{e},
Ronan Higgins\supit{d},
Kevin~Edwards\supit{f},
Adam~Brazier\supit{a}
\skiplinehalf
\supit{a}Department of Astronomy, Cornell University, Ithaca, NY, 14853, USA; \skiplinehalf
\supit{b}California Institute of Technology, 1200 E California Blvd, Pasadena, CA 91125, USA; \skiplinehalf
\supit{c}Argelander-Institut f\"{u}r Astronomie, Universit\"{a}t Bonn, Auf dem H\"{u}gel 71, 53121 Bonn, Germany; \skiplinehalf
\supit{d}KOSMA, I.\ Physikalisches Institut, Universit\"{a}t zu K\"{o}ln, Z\"{u}lpicher Str.\ 77, 50937 K\"{o}ln, Germany; \skiplinehalf
\supit{e}Department of Physics and Astronomy, University of British Columbia, 6224~Agricultural~Road, Vancouver, BC V6T~1Z1, Canada; \skiplinehalf
\supit{f}Department of Physics, University of Waterloo, Waterloo, ON N2L~3G1, Canada
}

\authorinfo{Further author information: (Send correspondence to T.J.)\\T.J.: E-mail: tjenness@cornell.edu}
\begin{document}
  \maketitle

%%%%%%%%%%%%%%%%%%%%%%%%%%%%%%%%%%%%%%%%%%%%%%%%%%%%%%%%%%%%%
\begin{abstract}
  CCAT will be a 25\,m diameter sub-millimeter telescope capable of
  operating in the 0.2 to 2.1\,mm wavelength range. It will be located
  at an altitude of 5600\,m on Cerro Chajnantor in northern Chile near
  the ALMA site. The anticipated first generation instruments include
  large
  format (60,000) kinetic inductance detector (KID) cameras, a large format heterodyne
  array and a direct detection
  multi-object spectrometer. The paper describes the architecture
  of the CCAT software and the development strategy.
\end{abstract}

%>>>> Include a list of keywords after the abstract

\keywords{Astronomy software, Facilities: CCAT, Software Development, Observatories, Submillimeter: general}

%%%%%%%%%%%%%%%%%%%%%%%%%%%%%%%%%%%%%%%%%%%%%%%%%%%%%%%%%%%%%
\section{Introduction}
\label{sec:intro}  % \label{} allows reference to this section

CCAT
\cite{2010SPIE.7733E..59S,2012SPIE.8444E..2MW,2013AAS...22115006G,P10_adassxxiii,2014HWTTU3Jenness}
is a 25\,m diameter sub-millimeter telescope to be built at an
altitude of 5600\,m on Cerro
Chajnantor in Chile near the ALMA site.
Operating at this altitude results in excellent transparency across
all observing bands from 350\micron\ to 2\,mm, including the potential
to observe at 200 um in the best weather
conditions. \cite{2014HWTTU3Jenness,2011RMxAC..41...87R}.

The CCAT project has identified four first generation
instruments to achieve its science goals. SWCam
\cite{2014SPIE9153-21,2013AAS...22115007S,2014SPIE9153-124} will
have of order 60,000 detectors operating mainly at
350\micron\ with additional detectors at 450, 850 and
2000\micron. CHAI \cite{GoldsmithCHAI2012}
will be a large format heterodyne array operating
simultaneously in two bands (two of 850, 600 and 350\micron) with at
least 32 elements per band, with the backend able to process spectra
with a bandwidth of at least 4\,GHz and 64k channels.  LWCam
\cite{2013AAS...22115008G} will be a dedicated long-wave camera operating
in 5-6 bands between 750\,\micron\ and 2.1\,mm with a long-wavelength
goal of 3.3\,mm. X-Spec \cite{2014SPIE9153-70,2013AAS...22115009B}
will be
a multi-object spectrometer with $\sim$\,100 beams on the sky, each
covering a frequency range of 190-520\,GHz in two bands simultaneously
with a resolving power of 400 -- 700.

\begin{figure}[t]
\begin{center}
\includegraphics[width=0.8\textwidth]{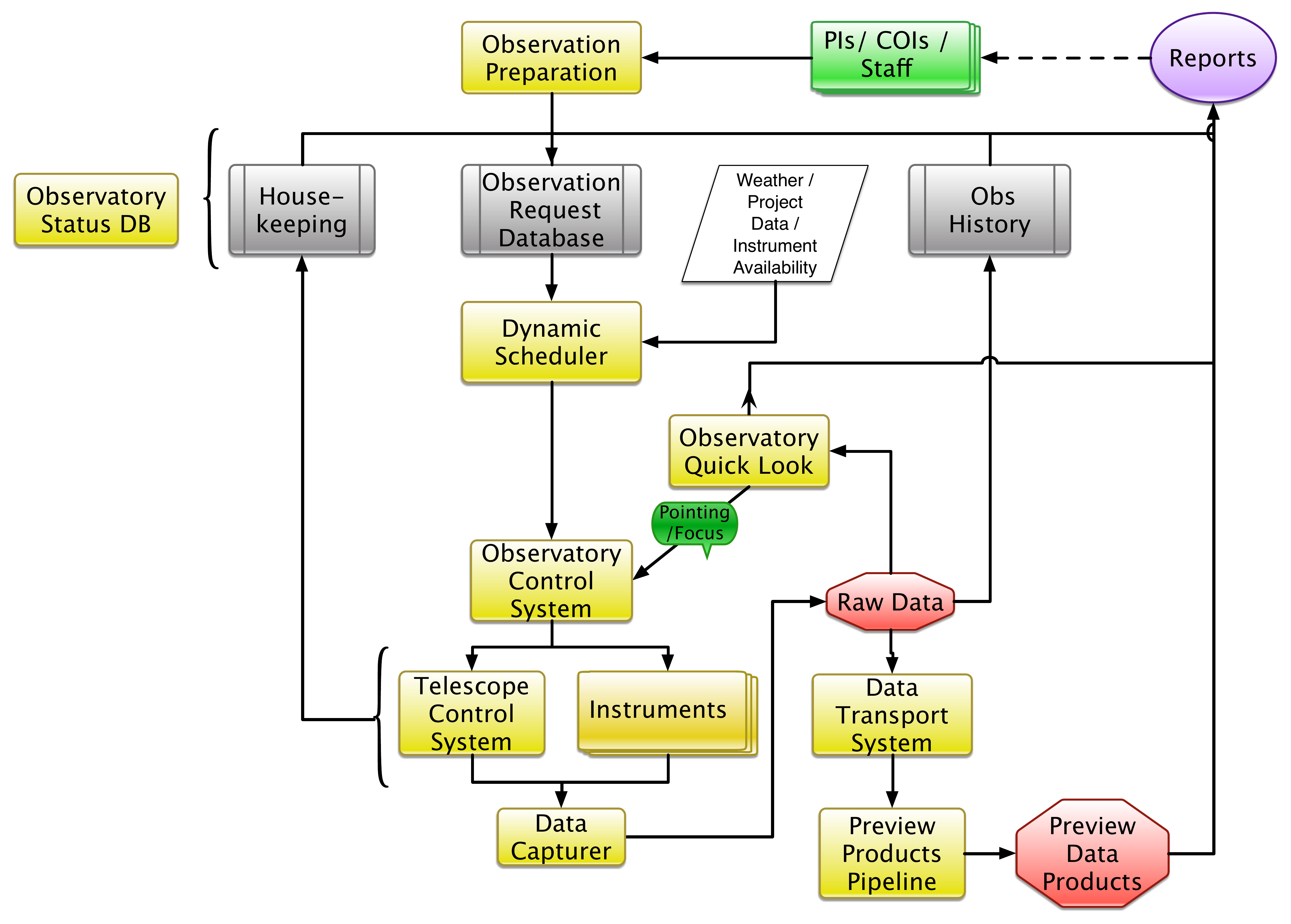}
\caption{Overview of CCAT software systems. Observations are prepared
  and stored in a database to be queried by the scheduler. The
  Observatory Control System
  executes observations and data flows from the instruments and
  telescope to the Data Capturer, which collates the output and makes
  the data files available to the Observatory Quick Look pipeline
  and data transport system. Preview data products are
  generated and made available to astronomers as part of reports on
  the status of their projects.}
\label{fig:SoftwareOverview}
\end{center}
\end{figure}

CCAT software development covers all phases of Observatory operations
(see Figure~\ref{fig:SoftwareOverview}),
including observation preparation, dynamic scheduling, observation
execution, data management and data reduction. In this paper we
present an overview on the software design and a report on current
status. The main drivers for the software design are:
\begin{enumerate}
\item Data rates that exceed a petabyte per year for the first
  generation instruments.
\item Remote operations of a telescope at an inhospitable site, with
  remote monitoring of observatory status by observers and engineers.
\end{enumerate}

The data rate for SWCam is expected to be of order 1~Gbps, assuming
some lossless compression is employed resulting in a variable bit
depth\cite{2014SPIE9153-124,2004SPIE.5498....1F}\footnote[1]{see for
  example the \texttt{slimdata} algorithm,
  \url{http://sourceforge.net/projects/slimdata/}}. The CHAI data rate
for the
baseline array design with 10\,Hz readout for on-the-fly mapping is
approximately 1.4\,Gbps. The estimated annual rate for CCAT, taking weather
statistics into account,\cite{2011RMxAC..41...87R} is of order 1.5
petabytes.

%% To make it easier to do distributed editing we break each section
%% into a standalone file. For submission these will have to be merged
%% into a single document.

\section{Observatory Control System}
\label{sec:ocs}

\begin{figure}[t]
\begin{center}
\includegraphics[height=3in]{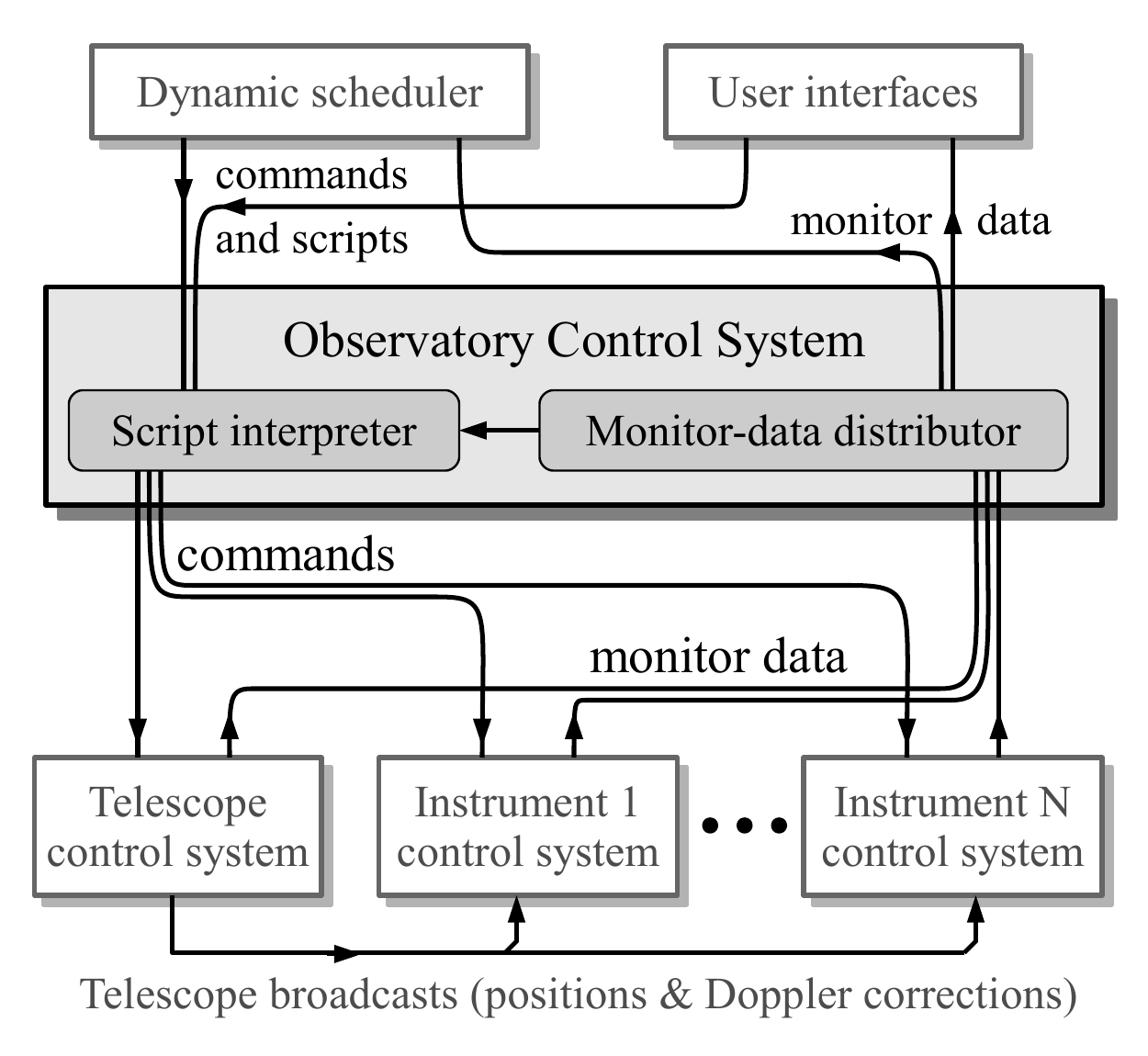}
\caption{The topology of the Observatory Control System (OCS). The OCS
contains a script interpreter, which sends commands to external
systems, and a component that receives monitoring values from external
systems and distributes them to user-interfaces, a dynamic scheduler
and the script interpreter.}
\label{ocs_topology_figure}
\end{center}
\end{figure}

The design of the Observatory Control System (OCS) for the CCAT
telescope is based on an approach used at a number of current and past
telescopes, such as the CBI\cite{2002PASP..114...83P},
QUIET\cite{2013ApJ...768....9B} and OVRO 40m\cite{2011ApJS..194...29R}
telescopes. Figure~\ref{ocs_topology_figure} shows the two major
parts of the OCS and how they interact with user interfaces,
the dynamic scheduler, the telescope control system (TCS), and various
instruments.

The OCS will be a minimal system that has no knowledge of the TCS, the
science instruments or any other systems until they connect to
it. When any system connects to the OCS, it will send the OCS the
declarations of the commands that it supports and the monitoring
values that it can supply. Thereafter, interactive users, the dynamic
scheduler and observing scripts will be able to send commands to these
systems and receive regular updates of dynamically-selected monitoring
values at 0.1\,s intervals. All systems that the OCS controls will
be treated as independent self-contained control-systems, each of
which will perform high level operations that are initiated by
commands sent to them by the OCS.

Coordination of the independent operations in the various instruments
and the TCS will be performed by scripts. The scripts will be written in a
custom high-level language, which has been designed for asynchronous
control with bounded latencies. The language has the ability to
quickly respond to monitoring feedback from any controlled
system, including the ability to cancel and preempt any ongoing
operation.

Communications between the OCS and controlled systems will occur on a
10\,Hz communication cycle. During each cycle, a block of commands and a
block of selected monitoring values will be exchanged between the OCS
and each controlled system. The sizes of these blocks are too large
for many industrial Ethernet protocols, and many of these protocols
don't allow dynamic reconfiguration for new instruments, or the
ability to send messages with varying contents. Rather than attempt to
design our own protocol, TCP/IP will be used.

Using TCP/IP for control takes some care.  TCP/IP is optimized for
continuous streams of data sent over large distances. It quickly
notices and retransmits dropped packets when the receiving TCP/IP
stack sees a gap in the stream of packet sequence-numbers, but this
only happens if more packets follow the dropped packets. A dropped
packet at the end of a short control message isn't noticed or
retransmitted for some time (0.2\,s between Linux hosts). A similar
problem occurs even when no packets are dropped, because of a well
known interaction between Nagle's algorithm and delayed
acknowledgments.\cite{2005Cheshire} To remedy both of these issues,
Nagle's algorithm is disabled by using the \texttt{TCP\_NODELAY}
configuration option, and the OCS communication scheme requires that
whenever a message is sent to a recipient, the recipient send back an
acknowledging reply. If this reply is not received within 25\,ms, the
sender transmits an extra message of a few packets, which simply asks
the recipient to echo this message back to the sender. The extra
packets in this echoed message reveal any gap in the stream of packets
to the receiving TCP/IP stack, which then initiates TCP/IP's
fast-retransmit scheme for the dropped packets.

The communication scheme described above has been tested on a cluster
of 11 computers.  One computer acted like the OCS server and the rest
acted like client instruments. The goal was to verify that the server
and each of the 10 client computers could repeatably exchange a
simulated 1\,kB-100\,kB monitoring message and ten simulated command
messages of 1\,kB size within a 0.1 second deadline. During a
contiguous period of 33 hours, the server and clients exchanged the
above messages at successive intervals of 0.1\,s, and all messages
were successfully delivered well within their 0.1\,s deadlines. This
was not true when the test was repeated without the countermeasures
discussed above. In that case many messages did miss their deadlines.

\section{Observation Preparation}
\label{sec:ot}

At CCAT there will be several ways of specifying an observation.
For experimentation and direct control of all instrument facilities an
approved observer can interact with the OCS directly and submit
scripts using the custom language detailed in section
\ref{sec:ocs}. They are in full control of observing and bypass the
dynamic scheduler.

For most observations,
observers will prepare minimum schedulable blocks
(MSBs) using the Observing Tool (OT). The OT can be used to prepare
MSBs containing one or more observations such that the technical
implementation details of a particular observing mode are hidden,
allowing the astronomer to focus on the science requirements for each
observation. The science specification of this MSB will be stored in a
database for later querying by the dynamic scheduler. It is only once
the scheduler has selected a particular MSB that the abstract view of
the observing will be translated into an OCS observing script ready
for execution.

Observers will also be able to supply an
OCS script of the core observing mode but uploaded to the observation
request database along with scheduling parameters and expected
duration. The dynamic scheduler will select the MSB as usual but there
will of course be no translation. Instead the scheduler will wrap the
supplied script with code that will firstly ensure that the resulting
data will be assigned to the correct observing project, and secondly,
ensure that the script is stopped if it takes significantly longer to
execute than was expected.

\subsection{Tool Selection}

A number of observing tools have been developed over the years
\cite{1997SPIE.3112..246W,2007AAS...210.1106F,2002ASPC..281..453F,1999AAS...19511907S,2005ESASP.560..895R,2012SPIE.8451E..1AB,2005AAS...20715905V}
and after careful consideration the project has adopted the JCMT-OT
\cite{2002ASPC..281..453F}, as used at the James Clerk Maxwell
Telescope (JCMT), to form the basis of the CCAT-OT for the initial
development phase. The JCMT-OT is the sub-millimeter ground-based
telescope version of the generic JAC-OT written by the Joint Astronomy
Centre that also supports the United Kingdom Infrared Telescope
(UKIRT)
\cite{2002SPIE.4844..321E,2000SPIE.4009..227B,2011tfa..confE..42J}.
It supports the HARP heterodyne receiver \cite{2009MNRAS.399.1026B}
and the SCUBA-2 bolometer array \cite{2013MNRAS.430.2513H}, which are
similar to CHAI and SWCam.

The JCMT-OT is written in Java and supports self-updating using
Java Web Start. Visualisation of the instruments on the sky is
provided by JSky \cite{1999Msngr..98...30D}. Proposal preparation is
not part of the JCMT-OT and we are considering using NorthStar
\cite{Northstar}, which would work with minor modifications.

\subsection{Modifications}

The JCMT-OT will require modifications in order to support CCAT.
In particular, code will need to be written to support the X-Spec
multi-object spectrometer so that the individual beams can be
positioned correctly.

MSB durations do not need to be extremely accurate, unlike for
satellite missions, and time estimates can be done locally in the tool
without requiring extensive server-side processing. An error of tens
of seconds in a ten minute observation is acceptable given
the use of a dynamic scheduler rather than a pre-determined night plan.

The atmospheric model used for the time estimators will have to be
modified for Chajnantor and the high frequency bands supported by
CHAI. We are also considering improving the
spectral editor capabilities of the JCMT-OT to more closely resemble the
functionality and look currently available in the ALMA-OT \cite{2013ASPC..475..373W}.

\begin{figure}[t]
\begin{center}
\includegraphics[width=\textwidth]{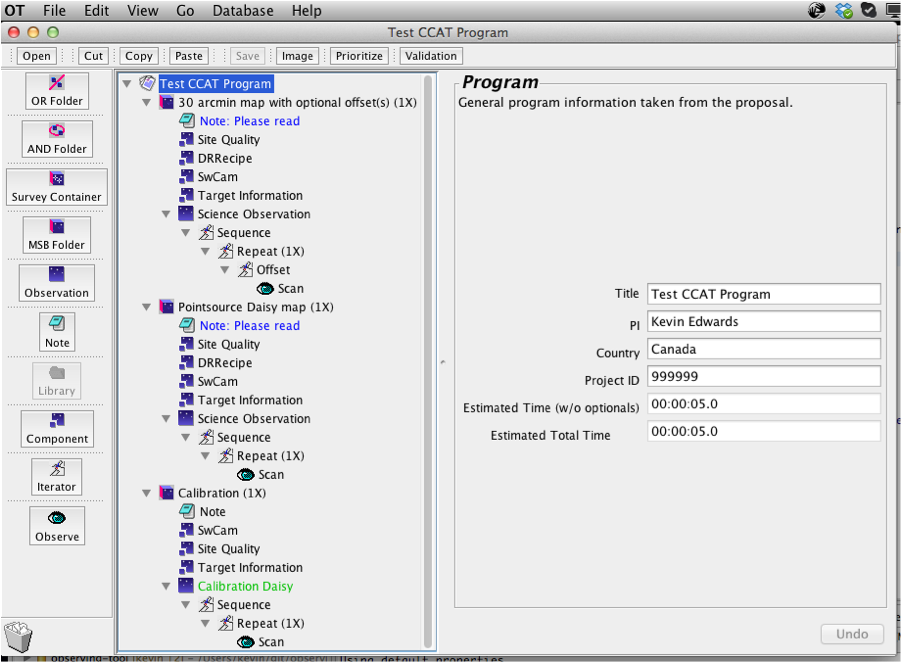}
\caption{The CCAT-OT prototype showing a science program that includes
  MSBs with an SWCam component.}
\label{fig:ot-swcam}
\end{center}
\end{figure}

The JCMT-OT source code is released under the GNU General Public License v2 for
the JCMT components, with the original Gemini code using the 3-clause
BSD license. The source code is available on a public
repository\footnote{\url{https://github.com/jac-h/observing-tool}}.
The code will need some
refactoring to meet the needs of CCAT users. The JCMT and UKIRT
components are integrated within the code base and will need to be
separated out. The code base is not supported by an accompanying set
of unit test code, so there will be some effort required to bring the
code up to more modern software development standards. These efforts
will largely go unseen by the general astronomer. The front-end
conversion of the tool to something that looks like a CCAT-OT will be
relatively straightforward as seen in
Figure~\ref{fig:ot-swcam} where we have already demonstrated the
ability to define an SWCam component.

\section{Raw Data Model}
\label{sec:datamodel}

The instrumental data rates are such that we have decided to use a
loosely-coupled distributed data acquisition system where each system controlled
by the OCS writes out a time-series independently of all other systems
but where synchronization is managed by accurate recording of time
stamps. It is up to the data reduction software to take the time
stamps and determine which sequences are related. Each system writes
out time-series using the same data model and a central data capturer
task (see \secref{sec:DM}) collates the individual components and creates a linker file
referencing them. The data capturer does not require highly
synchronized coordination of data writing between systems.

We have evaluated the existing data models
MBFITS\cite{2006A&A...454L..25M}, NDF\cite{jennessNDF,P91_adassxxiii} and
LOFAR\cite{2012ASPC..461..283A}. In view of CCAT's demands, each of these data models has its
advantages and disadvantages:
MBFITS keeps data from different systems in different files, in accordance with the envisaged
distributed data acquisition scheme; however, although MBFITS was designed and is being used for
continuum and spectral line arrays conceptually similar to CCAT's first-light instruments, the
data model is not flexible enough for new types of instruments like X-Spec.
In contrast, NDF gives data authors wide freedom to design specialized data models on top of the
general NDF model, and such data models for raw data (and data products) from continuum and spectral
line arrays are in use; however, since NDF lacks mechanisms to establish links between structures
in different files (such as the Hierarchical Grouping mechanism in FITS, or external links in HDF5),
such links can only be expressed by location of files in the file system and directory and file names.
The LOFAR Data Types (implemented in HDF5) are a family of related hierarchical data models for raw data and
data products for various LOFAR observing modes; they share common structures for common data and metadata
and allow specialized structures. However, the data models reflect the specific structure of the LOFAR array
and its observation modes too much to be used directly for a single-dish telescope with radically different
observing modes.

We have decided to develop a new data model based on HDF5
\cite{2011Folk:OHT:1966895.1966900}
as the low-level data format, but that shares
the merits of the evaluated data models.
During an observation, the data capturer, the TCS, and involved
instruments write their data to HDF5 files independently. The
set of these files forms a dataset that contains all data and
metadata of the involved systems during this observation.
In order to avoid excessive file sizes, bulk data from
the TCS and science instruments will be recorded in sequences of data
files which hold chunks of data for 30\,s each.

In order to form a single HDF5 hierarchy from the HDF5 structures in
the files of a dataset, HDF5's external links are used. The result is
a HDF5 hierarchy with basic observation-related metadata at the root
of the hierarchy, and TCS and instrument specific structures further
down the hierarchy.
Each OCS client system will write out structures in a standard way
such that the TCS component of a CHAI observation will be identical to
that of an SWCam observation. Furthermore, following the lead from
NDF, structure layouts will be re-used wherever possible when
designing the form of instrument-specific structures and, for example,
the time field in every time-series table will use the same name and
format to encourage code re-use and aid in cross-instrument
understanding. There is, however, no requirement for each instrument to
adopt data models that do not fit well with the
needs of the particular instrument.
This approach provides a good compromise between
a well-constrained model and one with sufficient flexibility to cope
with the specific needs of instruments.
Figure~\ref{fig:HdfHierarchy} illustrates the proposed top-level HDF5 hierarchy.

\begin{figure}[t]
\begin{center}
\small
\begin{quote}
\begin{verbatim}
  <ObsID>.h5                  Group                  Basic metadata
  |--OCS                      Group                  OCS metadata
  |--...                                             other static metadata
  |
  |--TCS                      Group (ext. link)      static TCS metadata
  |  |--TCS_00001             Group (ext. link)      TCS data
  |   --TCS_00002             Group (ext. link)      TCS data
  |
  |--SWCam350                 Group (ext. link)      static SWCam350 metadata
  |  |--SWCam350_00001        Group (ext. link)      SWCam350 data
  |   --SWCam350_00002        Group (ext. link)      SWCam350 data
  |
   --SWCam450                 Group (ext. link)      static SWCam450 metadata
     |--SWCam450_00001        Group (ext. link)      SWCam450 data
      --SWCam450_00002        Group (ext. link)      SWCam450 data

\end{verbatim}
\end{quote}
\caption{The HDF5 hierarchy of an observation with the 350\micron\ and
  450\micron\ sections of SWCam. The HDF5 root group and other groups
  directly below the root group (like the shown OCS
  group which holds OCS-related metadata) contain
  observation-related static metadata. TCS and instrument data and metadata
are stored in separate files, external links are used to establish a
single HDF5 hierarchy for all data in the dataset.}
\label{fig:HdfHierarchy}
\end{center}
\end{figure}

HDF5's external links rely on pathnames of the referenced files; since
absolute pathnames are not invariant when files are moved, only relative
pathnames are used. This requires that all files of a
dataset reside in a single directory tree. This can be achieved
with mounts and (file-system) symbolic links and it is also likely
that we will adopt the approach of using a distributed file system
such as GPFS. Figure~\ref{fig:DirectoryTree} sketches the
resulting directory structure.

\begin{figure}[t]
\begin{center}
\small
\begin{quote}
\begin{verbatim}
  <DataRoot>/<ObsID>/         Dir                   Root directory for observation dataset
  |--<ObsID>.h5               File                  Central linker file
  |
  |--TCS/                     Dir (symb. link)      Directory dedicated to TCS
  |  |--<ObsID>_00001.h5      File                  TCS data
  |   --<ObsID>_00002.h5      File                  TCS data
  |
  |--SWCam350/                Dir (symb. link)      Directory dedicated to SWCam350
  |  |--<ObsID>_00001.h5      File                  SWCam350 data
  |   --<ObsID>_00002.h5      File                  SWCam350 data
  |
   --SWCam450/                Dir (symb. link)      Directory dedicated to SWCam450
     |--<ObsID>_00001.h5      File                  SWCam450 data
      --<ObsID>_00002.h5      File                  SWCam450 data

\end{verbatim}
\end{quote}
\caption{Directories and files for the dataset sketched in
  Fig.~\ref{fig:HdfHierarchy}. The subdirectories with data from the
  TCS and instrument sections are mapped with mounts and symbolic links
into the central directory structure. Data from the TCS and
instruments is recorded in sequences of data files with 30\,s of data each.}
\label{fig:DirectoryTree}
\end{center}
\end{figure}

\section{Observatory Status Database}
\label{sec:osd}

The CCAT Observatory Status Database (OSD) stores and retrieves
information on all observatory operations and forms the central
``brain'' of the CCAT Observatory Data Management System. Comprising a
heterogeneous system of relational database and file system storage
with file readers, the OSD has a programmatic interface for input and
queries of operational CCAT data.

The OSD will have two geographically distinct components: the Live
OSD at the Observatory and the Legacy OSD hosted at a CCAT
Datacenter. Query access to both components will be through a single
interface and by default the querying party will not know which
component is serving their results.

\subsection{CCAT Observatory data management}
\label{sec:DM}

\begin{figure}[htb]
\begin{center}
\includegraphics[height=5in]{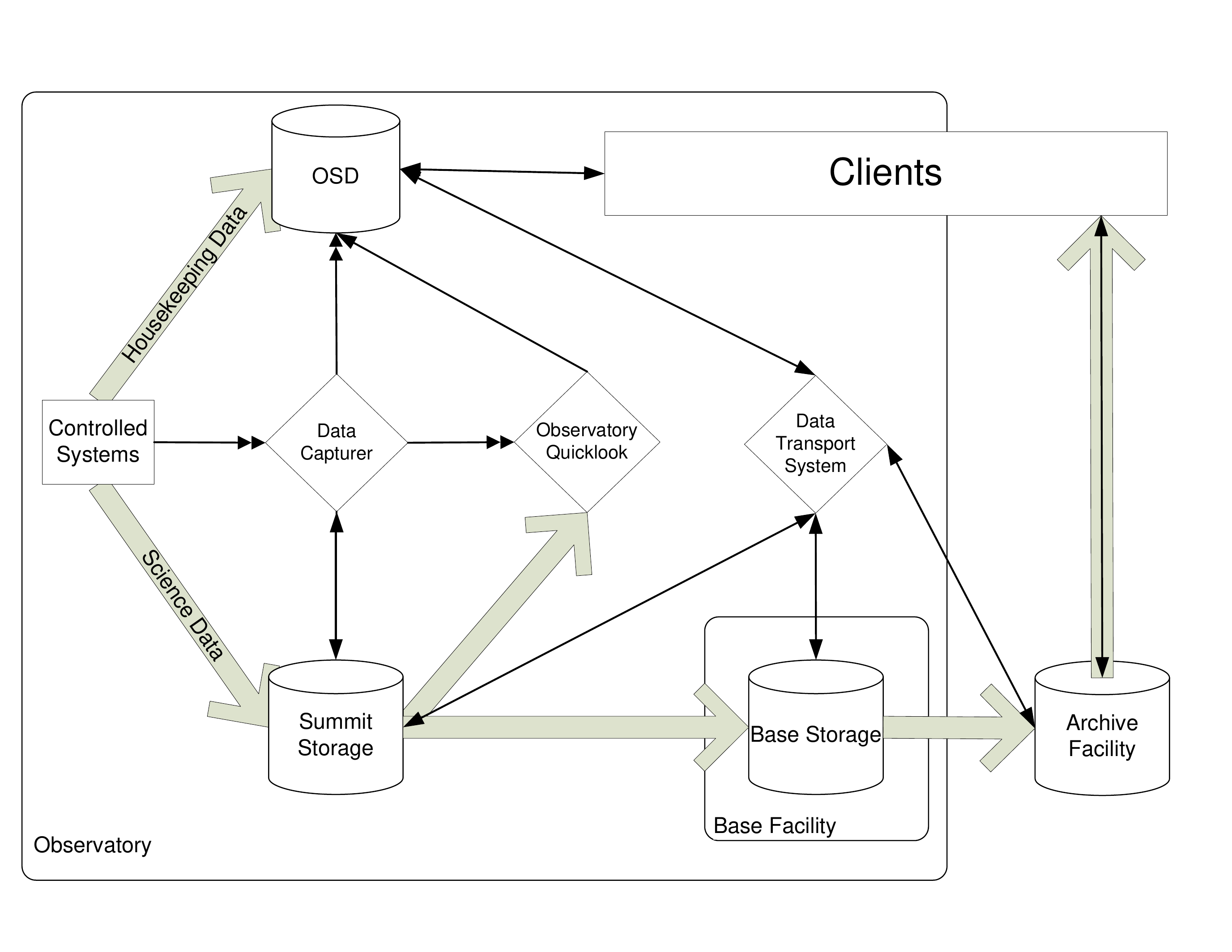}
\caption{CCAT Data Management Schema. Thick arrows represent bulk data
  flow and thinner arrows represent control and metadata
  communications, with double-headed arrows representing ``fast''
  connections with 1--3\,s latency.}
\label{fig:DM}
\end{center}
\end{figure}

Figure \ref{fig:DM} shows the overall data flow for CCAT data
management. Key components include:

\begin{itemize}
\item{Data Capturer: software system responsible for collating science
    data from the individual controlled systems, informing Obervatory Quicklook of the availability and
    location of the data and updating the OSD}
\item{Housekeeping Data (HK): all non-detector data from Controlled
    Systems}
\item{Data Transport System: Initiates network transfer and tracks
    network and physical transport of CCAT data files.}
\end{itemize}

\subsection{OSD Requirements}

In addition to standard data storage requirements such as those
relating to robustness and preservation of the integrity of
relationships between data, specific key requirements on the OSD
design include:

\begin{itemize}
\item{The OSD will have to consume HK data at a rate of $\sim$Mbits/s
    24 hours a day}
\item{The OSD shall have capacity to hold 30 days of HK data before
    transfer to the Legacy OSD for permanent storage.}
\item{The maximum allowed period after HK data are sent to the Live
    OSD to being available in query responses is 30 seconds}.
\end{itemize}

\subsection{OSD Prototype}

The OSD prototype is built in a development environment using the
following technologies:

\begin{itemize}
\item{Red Hat Enterprise Linux (RHEL) 6.5 for Controlled System and
    Data Storage machines including the file storage of HK data}
\item{Microsoft SQL Server 2012 running on Windows Server 2012}
\item{KVM for virtualization, running on an RHEL host}
\item{Python 3.4 for the OSD Application Programming Interface (API)}
\item{FreeTDS and \texttt{pyodbc} for database access (considering alternatives
    for ongoing development)}
\item{Stored Procedures for typical database operations, with
    ownership chaining, bound parameters and delineated permissions to
    control access to the database.}
\end{itemize}

The OSD prototype API consists of python modules to serve calls from
client code, which are validated against function specifications in
ancillary files and then executed, returning requested responses to
the calling code. The API is installed using \texttt{setuptools} onto the
client machine via a standard \texttt{setup.py} call and can be downloaded from
the Github repository in its entirety.

The OSD prototype is a collection of tables simply distinguishing
between data files, about which metadata are stored including the
location of the file, and ``data records'', data which are to be stored in a
SQL Server native type in database fields. A suite of stored
procedures written in static parametrized SQL ensure that insertions
of data and metadata preserve relationships between fields, which are
enforced by foreign keys and allow efficient querying of the OSD.

\section{Observation Management}

The management of observations is a critical component of a flexibly
scheduled telescope.\cite{2002ASPC..281..488E,2004SPIE.5493...24A,2014SPIE9149-51} In
general, Observations are defined by Scientists or by automated survey
definition tools and have to be tracked to ensure that the highest
priority observations are performed.

If someone is monitoring the observations there must be a way for
time-stamped commentary to be recorded to provide additional
information that is not available from the monitored computer
systems. This could involve a comment on data quality, the reason why
a particular instrument has been removed from the scheduler, or a
statement from an engineer regarding why an instrument warmed up. The
entry will contain the time the entry was made and also
the time for which the entry was relevant as the comment may be made
for some event that happened in the past. This system provides a
narrative log of events at the telescope and why decisions have been
made.

The data reduction pipelines running at the summit and base facility
will generate quality assurance parameters automatically and will
make this information available to the observing log.  Additionally it
shall be possible for people (for
example the current observer, instrument team, staff or collaborators)
to comment on a particular observation or a particular observation
block. This can be done during observing or later on after data have
been inspected more carefully.

It is important that scientists be able to inspect their data in near
real-time and if necessary modify their observing program to optimize
the science.
All the logging information, along with pipeline
products, quality assurance data, and monitoring data (such as weather
statistics) will be made available to the astronomers so they can make
informed decisions on data quality. There will also be a helpdesk
system to allow astronomers to ask questions about their data and
observing program.

\section{SWCam Data Reduction}
\label{sec:swcamdr}

Current ground-based submillimeter instruments, such as
SCUBA-2\cite{2013MNRAS.430.2513H}, SHARC-2\cite{2003SPIE.4855...73D}
and LABOCA,\cite{2009A&A...497..945S} have 100--1000s of
detectors. SWCam will have ${\sim}$60,000 detectors across four
wavelengths. Up to 48,000 detectors will be at 350\micron, the primary
wavelength for the instrument. Additionally, the SWCam KID detectors
will be sampled at ${\sim}$1000\,Hz (to a maximum rate of 1500\,Hz),
while the bolometer detectors on current instruments are sampled at
${\sim}$100\,Hz. This increase in sample rate is driven by the
combination of smaller beam sizes and faster telescope scan rates at CCAT compared to existing facilities.
The combined increase
in number of detectors and sample rate for SWCam compared to current
instruments results in a factor of 55 increase in data size over the
largest detectors today. It will be a
challenge to reduce these large datasets.

Map-making in the context of the next generation of submillimeter
cameras, such as SWCam, has been described
elsewhere.\cite{P14_adassxxiii} The above-mentioned
instruments all make use of iterative techniques to reduce common-mode
and correlated
noise.\cite{2013MNRAS.430.2545C,2008SPIE.7020E..45K,2012SPIE.8452E..1TS,2014SPIE9152-19}
These map-makers are based on similar algorithms; we focus on
SMURF,\cite{2013MNRAS.430.2545C,2013ascl.soft10007J} the SCUBA-2 map-maker, as it
was developed by the current team members, is highly configurable, is part
of the open source Starlink software collection \cite{P82_adassxxiii}\footnote{\url{http://www.starlink.ac.uk}}, and is in active development by the
Joint Astronomy Centre\cite{sc21}\footnote{\url{http://pipelinesandarchives.blogspot.com}}.
 Ref.~\citenum{P14_adassxxiii} estimates how the SMURF run time will scale from
SCUBA-2 to SWCam, and concludes that to keep up with data collection,
we will need either (i) several high-end machines running on
independent datasets or (ii) a cluster of machines running a
parallelized version of the map-maker. The parallel option has the
additional advantage that larger datasets (longer than about 15
minutes of observation) can be reduced without caching to disk, as the
data can be divided amongst the machines. In light of this, we have
investigated how SMURF could be modified to run on a
distributed-memory cluster.

\begin{figure}[t]
\begin{center}
\includegraphics[width=0.5\textwidth]{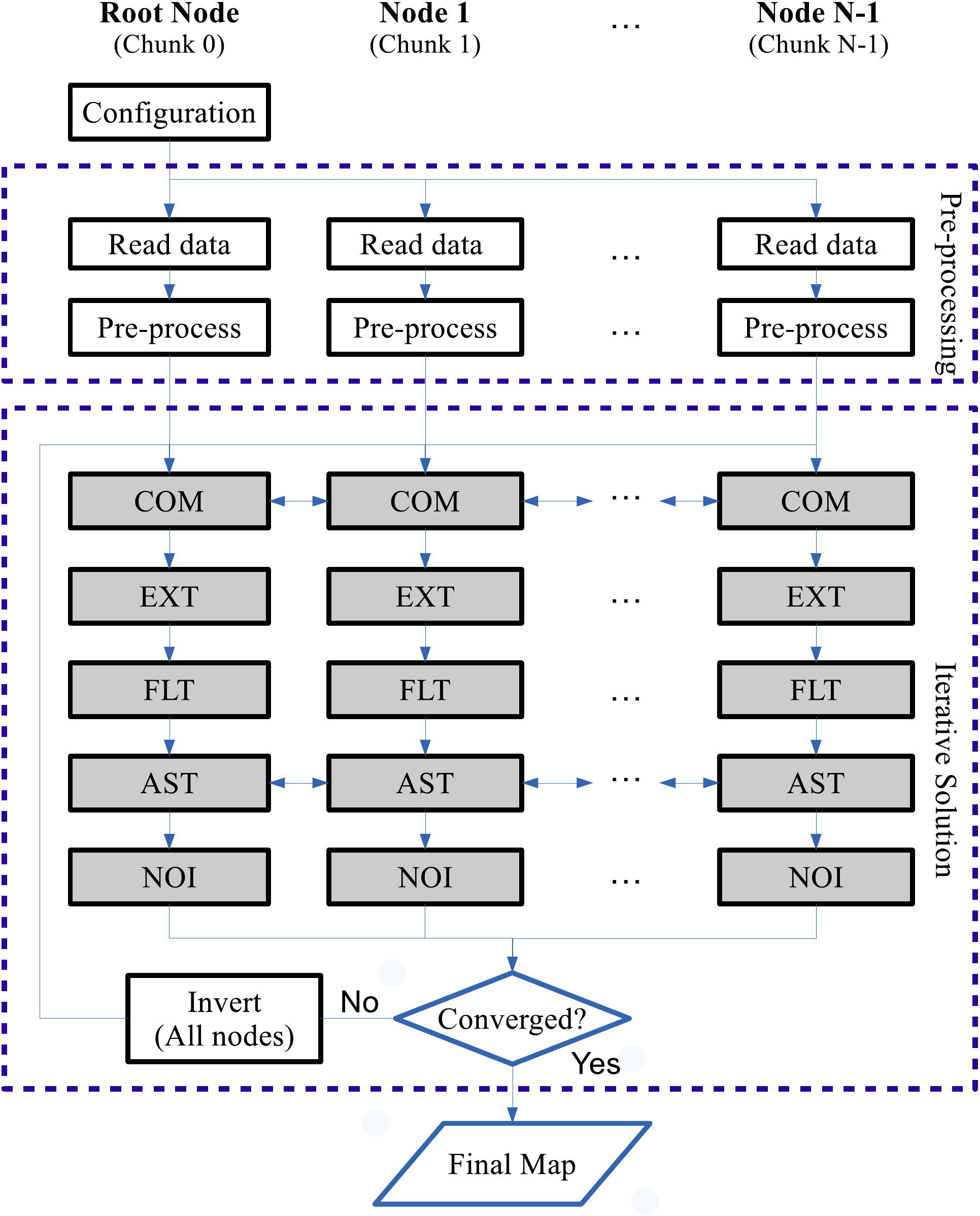}
\caption{A schematic for how distributed-memory parallelization of
  SMURF might work. The individual models are indicated by a three
  letter code: \texttt{COM} is the common-mode removal, \texttt{EXT}
  is atmospheric extinction correction, \texttt{FLT} is a Fourier
  filter, \texttt{AST} is the astronomical signal and \texttt{NOI} is
  a noise model. Horizontal arrows indicate modules where
  inter-node communication is required. The others are trivially
  parallelized. Figure reproduced from Ref.~\citenum{P14_adassxxiii}.}
\label{fig:ADASS_P14}
\end{center}
\end{figure}

Figure~\ref{fig:ADASS_P14} shows a schematic of how a
parallelized SMURF might be laid out. In this model, each process
manages a unique chunk of data, split so that each chunk contains the
full time streams for a number of detectors. The iterative algorithm
is modular, solving for a number of (user-specified) models
sequentially during each iteration. The program runs serially in each
process, with communications between processes occuring when
necessary. Some models, such as the filter model, which
applies a high-pass filter to each time stream in order to reduce
low-frequency detector and atmospheric noise, are trivially
parallelized, as each process can apply the filter to its chunk of
detectors, without the need to communicate with other processes. Other
models, such as the common-mode model, which removes a signal common
to all detectors at each time sample, require communication between
all processes. Due to this communication overhead, in addition to the
fact that the program will proceed at the rate of the slowest process,
it is not expected that the run time will scale as the inverse of the
number of processes.

A proof-of-concept implementation of the algorithm described
above has been
written\footnote{\url{https://github.com/CCATObservatory/mpi-mapmaker-test}}
to explore the scaling of run time with the number of processors for
a range of data sizes. The data sets used have 64, 256 and
4096 detectors, laid out in a square array, with the spacing adjusted
so that the detector array has the same angular size on the sky,
ensuring that the sky coverage is the same in all simulations. The simulated
data sets are created with 1800\,s of data and a sampling rate
of 1500 Hz.  The timing tests are run on Grex, a compute cluster that
is part of the WestGrid network.\footnote{\url{http://www.westgrid.ca}} Grex consists of a total of 316
nodes, each with two 6-core 2.66 GHz CPUs. The cluster features 24
nodes with 96 GB memory and the remaining 292 have 48 GB. For each
data set, we have run the map-maker using a range of number of
nodes. Where possible (namely the $N_\mathrm{det}=64$ and 256 data
sets), we start with one processor and increase the number by factors of
two until each process is operating on one detector ($N_\mathrm{proc}=64$
and 256, respectively). For the $N_\mathrm{ndet}$= 4096 data set, the
memory requirements prohibit using fewer than 8 processors (72~GB memory
required per process), and we stop at $N_\mathrm{proc}=1024$, a large
fraction of the available number of processors.

\begin{figure}[t]
\begin{center}
\includegraphics[width=\textwidth]{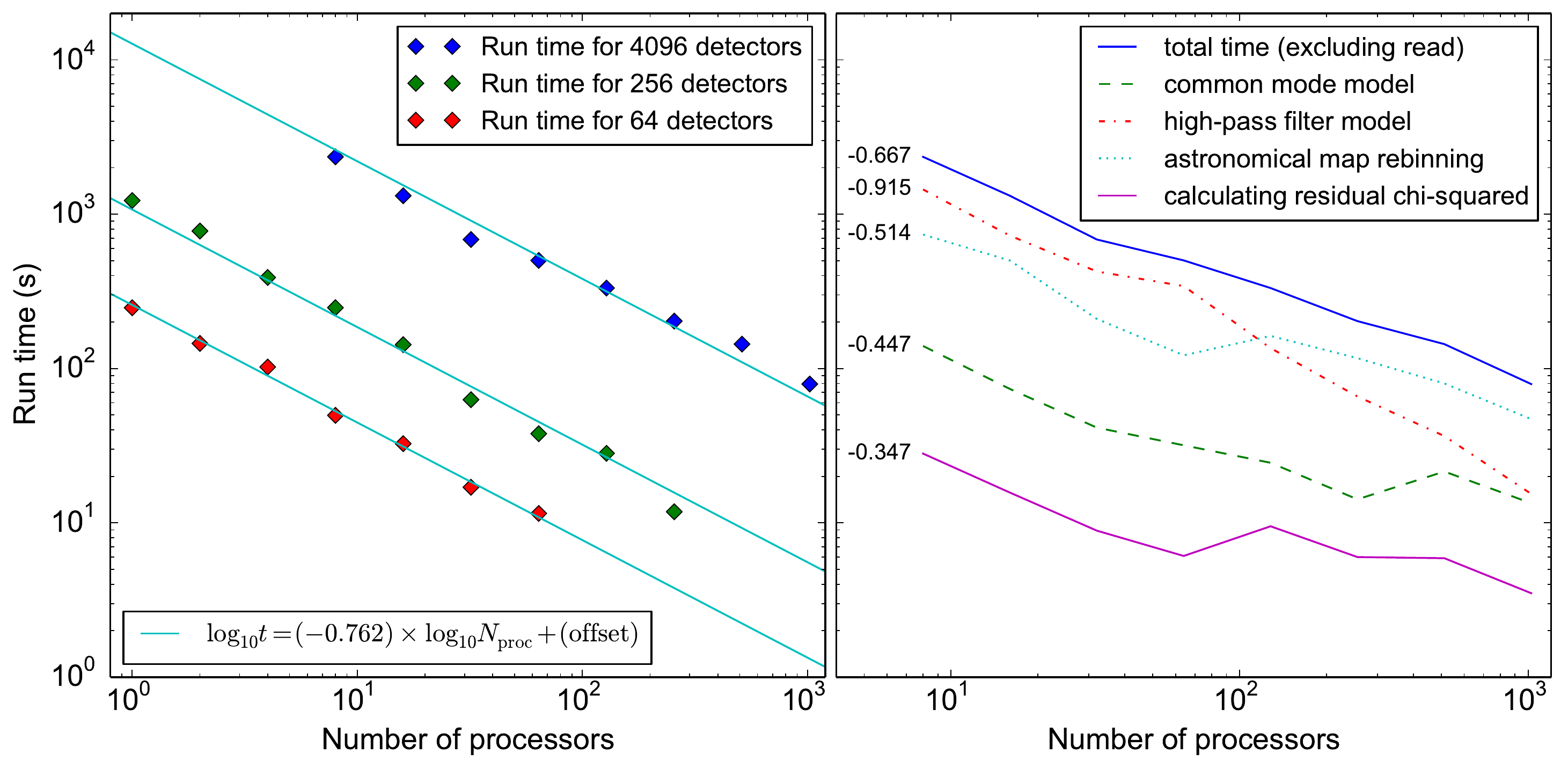}
\caption{Results of timing tests of SWCam prototype parallel
  map-maker. \textbf{Left panel:} Three timing tests for three
  detector counts (indicated by colored points). The input data
  consist of 1800\,s of data sampled at 1500\,Hz. All three tests are
  fit simultaneously with a common power-law index (solid lines),
  showing that a common slope is consistent with the
  data. \textbf{Right Panel:} Breakdown of subroutine run times for
  the $N_\mathrm{det}=4096$ data set. The best-fit power-law index for
  each component is indicated at the left edge of the associated
  curve.
\label{fig:swcam_dr_times}}
\end{center}
\end{figure}

The results of the timing tests are shown in the left panel of
Figure~\ref{fig:swcam_dr_times}. For each of the three datasets, run
time is plotted vs.\ number of processors used in the map-making
run. The datasets are simultaneously fit with a power law with common
power-law index, $t^{j} = A^{j} \times (N_\mathrm{proc})^{-\alpha}$,
where the index $j$ labels the three different datasets. The best-fit
index is $\alpha = 0.762$; this is shallower than perfect scaling
($\alpha = 1.0$), but that is to be expected due to communication
time overhead. It is encouraging, however, that the power-law relation
appears to continue to large $N_\mathrm{proc}$. While doubling the
number of available processors does not halve the run time, it does
significantly improve the run time, reducing it by about 40 per cent, and
additionally reduces the per-processor memory requirements (by about a
factor of two).

The right-hand panel of Figure~\ref{fig:swcam_dr_times} shows how the
run times of individual noise/signal models scale with the number of
processors. It shows the breakdown of run times for the
$N_\mathrm{det}=4096$ dataset, including the best-fit power-law
indices for each component (annotated at the left edge of each
line). We see that the high-pass filter model, which runs
independently in each process, not requiring communications between
processes, falls steeply, with a slope of nearly $\alpha = 1.0$. The
other models, the common-mode removal and map rebinning, require
communication between all processes, and thus do not scale as steeply
as $\alpha = 1.0$.

While the run time does not scale perfectly with the number of
processors used, the improvement is still
significant. Ref.~\citenum{P14_adassxxiii} states that 16 minutes of
SCUBA-2 data can be reduced on a dual quad-core CPU in 7 minutes using
33\,GB of memory. Using the data scaling factor of 55 from SCUBA-2 to
SWCam 350\micron, this same machine will take about six hours to
reduce 16 minutes of SWCam 350\micron\ data. Using the overall
best-fit power-law index of Figure \ref{fig:swcam_dr_times}, $\alpha=0.762$, a cluster of 50
of the above-described machines reduces the run time by a factor of 20
to $\sim 20$ minutes. The memory required is about 90\,GB per
node. Assuming that CCAT observes 12 hours per day, the parallel
algorithm described here will therefore be able to keep up with data
collection using mid-level hardware available today.

%%% This is probably not the best way to deal with footnotes. Is
%%% someone going to go through at the end and reset the counter at
%%% the appropriate spot?
\setcounter{footnote}{0}

\section{CHAI Data Reduction}
\label{sec:chaidr}

Even in its smallest configuration and readout rate CHAI will be
generating hundreds of thousands of spectra per night and for large
area on-the-fly mapping there may be millions. Reducing those
spectra requires automated pipelines capable of detecting bad spectra
and removing baseline artifacts with minimal input from a human operator.

The CHAI data reduction has to perform the reference subtraction,
including a correction for drifts, the frequency and intensity 
calibration, the flagging of known problems, such as bad channels or 
pixels, a baseline subtraction, and the evaluation of the resulting
data quality in terms of noise and the lack of unknown artifacts.
The pipeline
\cite{2008ASPC..394..565J,2013ASPC..475..341C,JennessACSISDR,2013ascl.soft10001J} for the
HARP instrument\cite{2009MNRAS.399.1026B} at the JCMT
is designed specifically for automated cube creation and we
are using it to investigate scaling and performance issues
from a 16-element instrument to larger arrays. We are testing the
pipeline on two data sets. The first data set comes from the
SMART\cite{2003SPIE.4855..322G,2005IAUS..235P.275K,2008stt..conf..488G} instrument on NANTEN2 providing direct
comparison with the current K\"{o}ln data reduction software written in CLASS which is part of the GILDAS data reduction package\cite{2013ascl.soft05010G}\footnote{\url{http://www.iram.fr/IRAMFR/GILDAS}}. The
second data set comes from commissioning data from the 64-element
Supercam\cite{2012SPIE.8452E..04K} on the Heinrich Hertz Submillimeter
Telescope (HHSMT). The Supercam data has eight times the detector
count of HARP although the channel count, 900 channels, is
nine times lower than the 8192 channels usually present in HARP
spectra. Both these data sets provide different tests of the pipeline
infrastructure and how the algorithm behaves as more detectors and
channels are used and will provide excellent feedback into the CHAI
pipeline design phase.

\section{Software Development}
\label{sec:swdev}

% Germany: Bonn (2), Cologne (2)
% Canada: UBC (1), UWaterloo (1)
% US: Caltech (2), Cornell (3), Tucson (2)

The CCAT software team is currently distributed over three countries,
seven institutions and twelve timezones and more institutions are expected to
contribute as the project enters the construction phase. Distributed
teams can result in difficulties in communication and the motivation of
isolated team members. Fast networks, ubiquitous webcam availability,
agile methodologies and advances in web site technologies continue to
aid distributed software development and within CCAT we have implemented
several approaches to maximize team effectiveness.

\subsection{Semi-annual face to face meetings}

There is still no replacement for face to face meetings to maximize
information transfer between team members. Full team meetings are held
every six months and are a critical aspect of team building. They
allow people to resolve misundertandings that built up
over the intervening months, as well as provide a social setting in the
evenings and during breaks to build up a rapport with other team
members that can not be achieved when you only know the person over
email in a professional context.

\subsection{Weekly ``standups''}

In the northern hemisphere summer there are twelve hours between
Hawaii and Germany and it is unreasonable to expect daily full team
meetings when participants in Hawaii have just woken up and
those in Germany are eating their evening meal. As a compromise we
have a full team video conference call each week to summarize progress
and report on any tasks that are being blocked. We have investigated a
number of different video conferencing technologies including Google
Hangout\footnote{\url{https://plus.google.com}},
GoToMeeting\footnote{\url{http://www.gotomeeting.com}} and
Zoom\footnote{\url{https://www.zoom.us}}. Each of these are capable of
screen sharing and ten video participants. They differ somewhat in
pricing strategies and the ability for people to call in from a
telephone. For example, Google Hangouts let you add people by calling
their number, GoToMeeting provides call in numbers for multiple
countries and Zoom provides a US toll-free number where the host must
pay a per-minute charge. Ideally people would call in using apps on
their smartphones and tablets if they are not using a computer but
during the transition from phone conferencing to video conferencing
there is still a need to support the telephone system.

\subsection{Team communication}

Mailing lists exist for each workpackage but email discussions can
become unwieldy as a topic is discussed over many days with many
levels of quoting. A mailing list is fine for a short broadcast to team members but we
are also considering collaborative instant messaging tools such as
Campfire,\footnote{\url{https://campfirenow.com}},
FlowDock\footnote{\url{https://www.flowdock.com}},
and HipChat.\footnote{\url{http://www.hipchat.com}} These tools allow general
conversations to occur throughout the day and lower the barrier for
asking quick questions to other team members. They do not, however, solve
the issues associated with debating larger topics over many days.

We are considering discussion tools such as
Discourse\footnote{\url{http://www.discourse.org}}. Fora have the best
potential for simplifying long form debates on a particular topic as
they allow quoting of particular paragraphs and responses to responses
whilst keeping the information in a single location and not spread
over a hundred emails. In some sense the code review features of
Github could easily serve a similar purpose although expecting people
to submit discussion topics to a git repository dedicated to this
purpose may be a step too far for people (developing
long-form documentation using Github is done regularly). Ideally it
should be possible to upload
gists\footnote{\url{https://gist.github.com}} and allow immediate
inline commentary but at present Github do not support this.

\subsection{Source Control and Collaboration}

%% Need to reset every 9 footnotes
\setcounter{footnote}{0}

Distributed revision control systems, such as \texttt{git}, that are
designed with distributed teams in mind are a huge aid to modern
software development. Branching in a repository is now seen as an
every day event rather than something that only the brave should
attempt.

We have looked at both Github and Atlassian's Bitbucket and have
decided to use Github since its collaboration tools are significantly
more powerful and easier to use. One key aspect is the integrated
wikis that are themselves hosted as git repositories and allow the use
of Markdown along with many other markup languages. We develop using
feature branches and make use of the code review features and issue
trackers provided by Github.

We have available up to 20 private git repositories that can be used for internal
document development and for early development of modules, but the
default is for all CCAT source code to be developed in public under a
3-clause BSD license in a similar approach to that taken by LSST
(albeit with a different, less restrictive,
license).\cite{2010SPIE.7740E..38A}\footnote{For details on the LSST
  software licensing policy see \url{https://dev.lsstcorp.org/trac/wiki/SWLicense}}

\subsection{Kanban boards}

The Kanban approach to software development\cite{2011Kniberg} is a
very popular agile technique but becomes difficult in a distributed
team when a physical Kanban board is being used. Online Kanban boards
are now available from many companies and currently we favor
Trello\footnote{\url{http://trello.com}} although the integration with
Github issues is not optimal.

%%%%%%%%%%%%%%%%%%%%%%%%%%%%%%%%%%%%%%%%%%%%%%%%%%%%%%%%%%%%%
%%%%% References %%%%%

\acknowledgments

The CCAT Submillimeter Observatory (CCAT) is owned and operated by a
consortium of universities and non-profit organizations located in the
United States, Canada and Germany. Specifically the CCAT Consortium is
comprised of: Cornell University, California Institute of Technology
(Caltech), University of Colorado at Boulder, University of Cologne,
University of Bonn, Dalhousie University, McGill University, McMaster
University, University of British Columbia, University of Calgary,
University of Toronto, University of Waterloo, University of Western
Ontario and Associated Universities, Incorporated.  The CCAT
Engineering Design Phase was partially supported by funding from the
National Science Foundation via AST-1118243. We thank William Peters
for making Supercam commissioning data available to us for testing.
We also thank Steve Padin, John Carpenter and Jeff Zivick for helpful comments on the manuscript.
The SWCam prototype map-maker makes use of facilities
provided by WestGrid and Compute Canada Calcul
Canada.\footnote{\url{http://www.computecanada.ca}}


\begin{thebibliography}{10}

\bibitem{2010SPIE.7733E..59S}
{Sebring}, T., ``{The Cornell Caltech Atacama Telescope: progress and plans
  2010},'' in [{\em Ground-based and Airborne Telescopes
  III}{\nolinebreak\hspace{0.1em}]},  Stepp, L.~M., Gilmozzi, R., and Hall,
  H.~J., eds., {\em Proc.\ SPIE} {\bf 7733},  77331X (2010).

\bibitem{2012SPIE.8444E..2MW}
{Woody}, D., Padin, S., Chauvin, E., et~al., ``{The CCAT 25m diameter
  submillimeter-wave telescope},'' in [{\em Ground-based and Airborne
  Telescopes IV}{\nolinebreak\hspace{0.1em}]},  Stepp, L.~M., Gilmozzi, R., and
  Hall, H.~J., eds., {\em Proc.\ SPIE} {\bf 8444},  84442M (2012).

\bibitem{2013AAS...22115006G}
{Glenn}, J. et~al., ``{The CCAT Telescope},'' in [{\em American Astronomical
  Society Meeting}{\nolinebreak\hspace{0.1em}]},  {\em American Astronomical
  Society Meeting Abstracts} {\bf 221},  \#150.06 (2013).

\bibitem{P10_adassxxiii}
Jenness, T., Brazier, A., Edwards, K., et~al., ``{The CCAT software System},''
  in [{\em Astronomical Data Analysis Software and Systems
  XXIII}{\nolinebreak\hspace{0.1em}]},  Manset, N. and Forshay, P., eds., {\em
  \aspconf} {\bf 485},  49, ASP, San Francisco (2014).
\newblock arXiv:1401.8280.

\bibitem{2014HWTTU3Jenness}
Jenness, T., ``{Transient Alert Follow-up Planned for CCAT},'' in [{\em
  {Hot-Wiring the Transient Universe 3}}{\nolinebreak\hspace{0.1em}]},
  Wozniak, P., Graham, M., and Mahabal, A., eds. (2014).
\newblock arXiv:1402.1202.

\bibitem{2011RMxAC..41...87R}
{Radford}, S.~J.~E., ``{Observing Conditions for Submillimeter Astronomy},'' in
  [{\em Astronomical Site Testing Data in Chile}{\nolinebreak\hspace{0.1em}]},
  {\em Revista Mexicana de Astronomia y Astrofisica Conference Series} {\bf
  41},  87--90 (2011).

\bibitem{2014SPIE9153-21}
Stacey, G.~J., Parshley, S., Nikola, T., et~al., ``{SWCam: the short wavelength
  camera for the CCAT observatory},'' in [{\em Millimeter, Submillimeter, and
  Far-Infrared Detectors and Instrumentation for Astronomy
  VII}{\nolinebreak\hspace{0.1em}]},  Holland, W.~S. and Zmuidzinas, J., eds.,
  {\em Proc.\ SPIE} {\bf 9153},  915321 (2014).

\bibitem{2013AAS...22115007S}
{Stacey}, G.~J., Parshley, S., Nikola, T., et~al., ``{The Design of the Short
  Wavelength Camera for the CCAT Telescope},'' in [{\em American Astronomical
  Society Meeting}{\nolinebreak\hspace{0.1em}]},  {\em American Astronomical
  Society Meeting Abstracts} {\bf 221},  \#150.07 (2013).

\bibitem{2014SPIE9153-124}
Rajagopalan, G., Kovacs, A., Monroe, R.~M., et~al., ``{Readout electronics for
  the CCAT observatory's instruments at first light and beyond},'' in [{\em
  Millimeter, Submillimeter, and Far-Infrared Detectors and Instrumentation for
  Astronomy VII}{\nolinebreak\hspace{0.1em}]},  Holland, W.~S. and Zmuidzinas,
  J., eds., {\em Proc.\ SPIE} {\bf 9153},  9153124 (2014).

\bibitem{GoldsmithCHAI2012}
Goldsmith, P. et~al., ``{Studying the Formation and Development of Molecular
  Clouds: with the CCAT Heterodyne Array Instrument (CHAI)}.''
  \url{http://trs-new.jpl.nasa.gov/dspace/bitstream/2014/42645/1/12-2032_A1b.pdf}
  (2012).

\bibitem{2013AAS...22115008G}
{Golwala}, S.~R. et~al., ``{The Design and Science Goals of LWCam, the CCAT
  Long-Wavelength Imager},'' in [{\em American Astronomical Society
  Meeting}{\nolinebreak\hspace{0.1em}]},  {\em American Astronomical Society
  Meeting Abstracts} {\bf 221},  \#150.08 (2013).

\bibitem{2014SPIE9153-70}
Bradford, C.~M., Hailey-Dunsheath, S., Shirokoff, E.~D., Hollister, M.~I.,
  McKenney, C.~M., Chapman, S., Tikhomirov, A., and Nikola, T., ``{X-Spec: a
  multi-object trans-millimeter-wave spectrometer for CCAT},'' in [{\em
  Millimeter, Submillimeter, and Far-Infrared Detectors and Instrumentation for
  Astronomy VII}{\nolinebreak\hspace{0.1em}]},  Holland, W.~S. and Zmuidzinas,
  J., eds., {\em Proc.\ SPIE} {\bf 9153},  915370 (2014).

\bibitem{2013AAS...22115009B}
{Bradford}, C., Hailey-Dunsheath, S., Shirokoff, E., et~al., ``{X-Spec: A
  Multi-Object Wideband Survey Spectrograph for CCAT},'' in [{\em American
  Astronomical Society Meeting}{\nolinebreak\hspace{0.1em}]},  {\em American
  Astronomical Society Meeting Abstracts} {\bf 221},  \#150.09 (2013).

\bibitem{2004SPIE.5498....1F}
{Fowler}, J.~W., ``{The Atacama Cosmology Telescope Project},'' in [{\em
  Millimeter and Submillimeter Detectors for Astronomy
  II}{\nolinebreak\hspace{0.1em}]},  Zmuidzinas, J., Holland, W.~S., and
  Withington, S., eds., {\em Proc.\ SPIE} {\bf 5498},  1--10 (2004).

\bibitem{2002PASP..114...83P}
{Padin}, S., {Shepherd}, M.~C., {Cartwright}, J.~K., et~al., ``{The Cosmic
  Background Imager},'' {\em \pasp}~{\bf 114},  83--97 (2002).

\bibitem{2013ApJ...768....9B}
{Bischoff}, C. et~al., ``{The Q/U Imaging ExperimenT Instrument},'' {\em
  \apj}~{\bf 768},  9 (2013).

\bibitem{2011ApJS..194...29R}
{Richards}, J.~L., {Max-Moerbeck}, W., {Pavlidou}, V., et~al., ``{Blazars in
  the Fermi Era: The OVRO 40 m Telescope Monitoring Program},'' {\em
  \apjs}~{\bf 194},  29 (2011).

\bibitem{2005Cheshire}
Cheshire, S., ``{TCP Performance problems caused by interaction between
  Nagle’s Algorithm and Delayed ACK}.''
  \url{http://www.stuartcheshire.org/papers/NagleDelayedAck/} (2005).

\bibitem{1997SPIE.3112..246W}
{Wampler}, S., {Gillies}, K.~K., {Puxley}, P.~J., and {Walker}, S., ``{Science
  planning for the Gemini 8-m telescopes},'' in [{\em Telescope Control Systems
  II}{\nolinebreak\hspace{0.1em}]},  {Lewis}, H., ed., {\em Proc.\ SPIE} {\bf
  3112},  246--253 (1997).

\bibitem{2007AAS...210.1106F}
{Frayer}, D.~T. et~al., ``{Tools for the Herschel Space Observatory:
  Observation Planning and Data Processing},'' in [{\em American Astronomical
  Society Meeting Abstracts \#210}{\nolinebreak\hspace{0.1em}]},  {\em Bulletin
  of the American Astronomical Society} {\bf 39},  108 (2007).

\bibitem{2002ASPC..281..453F}
{Folger}, M., {Bridger}, A., {Dent}, B., {Kelly}, D., {Adamson}, A.,
  {Economou}, F., {Hirst}, P., and {Jenness}, T., ``{A New Observing Tool for
  the James Clerk Maxwell Telescope},'' in [{\em Astronomical Data Analysis
  Software and Systems XI}{\nolinebreak\hspace{0.1em}]},  {Bohlender}, D.~A.,
  {Durand}, D., and {Handley}, T.~H., eds., {\em \aspconf} {\bf 281},  453
  (2002).

\bibitem{1999AAS...19511907S}
{Storrie-Lombardi}, L.~J. and {SSC Team}, ``{Planning Observations for a SIRTF
  Legacy Science Proposal},'' in [{\em American Astronomical Society Meeting
  Abstracts}{\nolinebreak\hspace{0.1em}]},  {\em Bulletin of the American
  Astronomical Society} {\bf 31},  \#119.07 (1999).

\bibitem{2005ESASP.560..895R}
{Rebull}, L.~M. and {SSC Observer Support Team}, ``{Planning your observations
  with the Spitzer Space Telescope},'' in [{\em 13th Cambridge Workshop on Cool
  Stars, Stellar Systems and the Sun}{\nolinebreak\hspace{0.1em}]},  {Favata},
  F., {Hussain}, G.~A.~J., and {Battrick}, B., eds., {\em ESA Special
  Publication} {\bf 560},  895 (2005).

\bibitem{2012SPIE.8451E..1AB}
{Bridger}, A., {Williams}, S., {McLay}, S., {Yatagai}, H., {Schilling}, M.,
  {Biggs}, A., {Tobar}, R., and {Warmels}, R.~H., ``{The ALMA OT in early
  science: supporting multiple customers},'' in [{\em Software and
  Cyberinfrastructure for Astronomy II}{\nolinebreak\hspace{0.1em}]},
  Radziwill, N.~M. and Chiozzi, G., eds., {\em Proc. SPIE} {\bf 8451},  84511A
  (2012).

\bibitem{2005AAS...20715905V}
{Vacca}, W.~D. and {Sandell}, G., ``{US SOFIA General Investigator Program: How
  to Propose; How to Plan},'' in [{\em American Astronomical Society Meeting
  Abstracts}{\nolinebreak\hspace{0.1em}]},  {\em Bulletin of the American
  Astronomical Society} {\bf 37},  \#159.05 (2005).

\bibitem{2002SPIE.4844..321E}
{Economou}, F., {Jenness}, T., and {Rees}, N.~P., ``{Sharing code and support
  between heterogeneous telescopes: the UKIRT and JCMT joint software
  projects},'' in [{\em Observatory Operations to Optimize Scientific Return
  III}{\nolinebreak\hspace{0.1em}]},  {Quinn}, P.~J., ed., {\em Proc.\ SPIE}
  {\bf 4844},  321--330 (2002).

\bibitem{2000SPIE.4009..227B}
{Bridger}, A., {Wright}, G.~S., {Economou}, F., et~al., ``{ORAC: a modern
  observing system for UKIRT},'' in [{\em Advanced Telescope and
  Instrumentation Control Software}{\nolinebreak\hspace{0.1em}]},  {Lewis}, H.,
  ed., {\em Proc.\ SPIE} {\bf 4009},  227--238 (2000).

\bibitem{2011tfa..confE..42J}
{Jenness}, T. and {Economou}, F., ``{Data Management at the UKIRT and JCMT},''
  in [{\em Telescopes from Afar}{\nolinebreak\hspace{0.1em}]},  {Gajadhar}, S.,
  {Walawender}, J., {Genet}, R., {Veillet}, C., {Adamson}, A., {Martinez}, J.,
  {Melnik}, J., {Jenness}, T., and {Manset}, N., eds.,  42 (2011).
\newblock arXiv:1111.5855.

\bibitem{2009MNRAS.399.1026B}
{Buckle}, J.~V., {Hills}, R.~E., {Smith}, H., et~al., ``{HARP/ACSIS: a
  submillimetre spectral imaging system on the James Clerk Maxwell
  Telescope},'' {\em \mnras}~{\bf 399},  1026--1043 (2009).

\bibitem{2013MNRAS.430.2513H}
{Holland}, W.~S., {Bintley}, D., {Chapin}, E.~L., et~al., ``{SCUBA-2: the 10
  000 pixel bolometer camera on the James Clerk Maxwell Telescope},'' {\em
  \mnras}~{\bf 430},  2513--2533 (2013).

\bibitem{1999Msngr..98...30D}
{Dolensky}, M., {Albrecht}, M., {Albrecht}, R., et~al., ``{Java for astronomy:
  software development at ESO/ST-ECF.},'' {\em The Messenger}~{\bf 98},  30--32
  (1999).

\bibitem{Northstar}
Smit, A. and Holties, H., ``{NorthStar --- The RadioNet Common Proposal
  Tool}.'' AstroGrid Science Workshop and RadioNet/AstroGrid workshop for radio
  data providers, Oxford University (2006).
\newblock
  \url{http://wiki.astrogrid.org/pub/Astrogrid/RadioAgenda/NSAstroGrid08122006.pdf}.

\bibitem{2013ASPC..475..373W}
{Williams}, S. and {Bridger}, A., ``{Spectral Line Selection in the ALMA
  Observing Tool},'' in [{\em Astronomical Data Analysis Software and Systems
  XXII}{\nolinebreak\hspace{0.1em}]},  {Friedel}, D.~N., ed., {\em \aspconf}
  {\bf 475},  373 (2013).

\bibitem{2006A&A...454L..25M}
{Muders}, D., {Hafok}, H., {Wyrowski}, F., et~al., ``{APECS - the Atacama
  pathfinder experiment control system},'' {\em \aap}~{\bf 454},  L25--L28
  (2006).

\bibitem{jennessNDF}
Jenness, T., Berry, D.~S., Currie, M.~J., et~al., ``{Learning from 25 years of
  the extensible \emph{N}-Dimensional Data Format},'' {\em Astron.\ Comp.}~{\bf
  submitted} (2014).

\bibitem{P91_adassxxiii}
Economou, F., Jenness, T., Currie, M.~J., and Berry, D.~S., ``{Advantages of
  Extensible Self-described Data Formats: Lessons Learned from NDF},'' in [{\em
  Astronomical Data Analysis Software and Systems
  XXIII}{\nolinebreak\hspace{0.1em}]},  Manset, N. and Forshay, P., eds., {\em
  \aspconf} {\bf 485},  355 (2014).

\bibitem{2012ASPC..461..283A}
{Alexov}, A., {Schellart}, P., {ter Veen}, S., et~al., ``{Status of LOFAR Data
  in HDF5 Format},'' in [{\em Astronomical Data Analysis Software and Systems
  XXI}{\nolinebreak\hspace{0.1em}]},  {Ballester}, P., {Egret}, D., and
  {Lorente}, N.~P.~F., eds., {\em \aspconf} {\bf 461},  283 (2012).

\bibitem{2011Folk:OHT:1966895.1966900}
Folk, M., Heber, G., Koziol, Q., Pourmal, E., and Robinson, D., ``{An Overview
  of the HDF5 Technology Suite and Its Applications},'' in [{\em Proceedings of
  the EDBT/ICDT 2011 Workshop on Array Databases}{\nolinebreak\hspace{0.1em}]},
   {\em AD '11},  36--47, ACM (2011).

\bibitem{2002ASPC..281..488E}
{Economou}, F., {Jenness}, T., {Tilanus}, R.~P.~J., {Hirst}, P., {Adamson},
  A.~J., {Rippa}, M., {Delorey}, K.~K., and {Isaak}, K.~G., ``{Flexible
  Software for Flexible Scheduling},'' in [{\em Astronomical Data Analysis
  Software and Systems XI}{\nolinebreak\hspace{0.1em}]},  {Bohlender}, D.~A.,
  {Durand}, D., and {Handley}, T.~H., eds., {\em \aspconf} {\bf 281},  488
  (2002).

\bibitem{2004SPIE.5493...24A}
{Adamson}, A.~J., {Tilanus}, R.~P., {Buckle}, J., {Davis}, G.~R., {Economou},
  F., {Jenness}, T., and {Delorey}, K., ``{Science returns of flexible
  scheduling on UKIRT and the JCMT},'' in [{\em Optimizing Scientific Return
  for Astronomy through Information Technologies}{\nolinebreak\hspace{0.1em}]},
   {Quinn}, P.~J. and {Bridger}, A., eds., {\em Proc.\ SPIE} {\bf 5493},
  24--32 (2004).

\bibitem{2014SPIE9149-51}
Dempsey, J.~T., Jenness, T., Economou, F., et~al., ``{Setting the standard: 25
  years of operating the JCMT},'' in [{\em Observatory Operations: Strategies,
  Processes, and Systems V}{\nolinebreak\hspace{0.1em}]},  Peck, A.~B., Benn,
  C.~R., and Seaman, R., eds., {\em Proc.\ SPIE} {\bf 9149},  914951 (2014).

\bibitem{2003SPIE.4855...73D}
{Dowell}, C.~D., {Allen}, C.~A., {Babu}, R.~S., et~al., ``{SHARC II: a Caltech
  submillimeter observatory facility camera with 384 pixels},'' in [{\em
  Millimeter and Submillimeter Detectors for
  Astronomy}{\nolinebreak\hspace{0.1em}]},  {Phillips}, T.~G. and {Zmuidzinas},
  J., eds., {\em Proc.\ SPIE} {\bf 4855},  73--87 (2003).

\bibitem{2009A&A...497..945S}
{Siringo}, G., {Kreysa}, E., {Kov{\'a}cs}, A., et~al., ``{The Large APEX
  BOlometer CAmera LABOCA},'' {\em \aap}~{\bf 497},  945--962 (2009).

\bibitem{P14_adassxxiii}
{Marsden}, G., {Brazier}, A., {Jenness}, T., {Sayers}, J., and {Scott}, D.,
  ``{Map-making for the Next Generation of Ground-based Submillimeter
  Instruments},'' in [{\em Astronomical Data Analysis Software and Systems
  XXIII}{\nolinebreak\hspace{0.1em}]},  Manset, N. and Forshay, P., eds., {\em
  \aspconf} {\bf 485},  399, ASP, San Francisco (2014).
\newblock arXiv:1405.0482.

\bibitem{2013MNRAS.430.2545C}
{Chapin}, E.~L., {Berry}, D.~S., {Gibb}, A.~G., {Jenness}, T., {Scott}, D.,
  {Tilanus}, R.~P.~J., {Economou}, F., and {Holland}, W.~S., ``{SCUBA-2:
  iterative map-making with the Sub-Millimetre User Reduction Facility},'' {\em
  \mnras}~{\bf 430},  2545--2573 (2013).

\bibitem{2008SPIE.7020E..45K}
{Kov{\'a}cs}, A., ``{CRUSH: fast and scalable data reduction for imaging
  arrays},'' in [{\em Millimeter and Submillimeter Detectors and
  Instrumentation for Astronomy IV}{\nolinebreak\hspace{0.1em}]},  Duncan,
  W.~D., Holland, W.~S., and Withington, S., eds., {\em Proc. SPIE} {\bf 7020},
   70201S (2008).

\bibitem{2012SPIE.8452E..1TS}
{Schuller}, F., ``{BoA: a versatile software for bolometer data reduction},''
  in [{\em Millimeter, Submillimeter, and Far-Infrared Detectors and
  Instrumentation for Astronomy VI}{\nolinebreak\hspace{0.1em}]},  Holland,
  W.~S., ed., {\em Proc. SPIE} {\bf 8452},  84521T (2012).

\bibitem{2014SPIE9152-19}
Kov{\'a}cs, A., ``{CRUSH: data reduction and imaging for future (sub)millimeter
  arrays},'' in [{\em Software and Cyberinfrastructure for Astronomy
  III}{\nolinebreak\hspace{0.1em}]},  Chiozzi, G. and Radziwill, N.~M., eds.,
  {\em Proc.\ SPIE} {\bf 9152},  915219 (2014).

\bibitem{2013ascl.soft10007J}
{Jenness}, T., {Chapin}, E.~L., {Berry}, D.~S., {Gibb}, A.~G., {Tilanus},
  R.~P.~J., {Balfour}, J., {Tilanus}, V., and {Currie}, M.~J., ``{SMURF:
  SubMillimeter User Reduction Facility},'' (2013).
\newblock Astrophysics Source Code Library \ascl{1310.007}.

\bibitem{P82_adassxxiii}
Currie, M.~J., Berry, D.~S., Jenness, T., Gibb, A.~G., Bell, G.~S., and Draper,
  P.~W., ``Starlink in 2013,'' in [{\em Astronomical Data Analysis Software and
  Systems XXIII}{\nolinebreak\hspace{0.1em}]},  Manset, N. and Forshay, P.,
  eds., {\em \aspconf} {\bf 485},  391 (2014).

\bibitem{sc21}
Thomas, H.~S., ``{The SCUBA-2 Data Reduction Cookbook}.'' Starlink Cookbook 21,
  Joint Astronomy Centre (2013).

\bibitem{2008ASPC..394..565J}
{Jenness}, T., {Cavanagh}, B., {Economou}, F., and {Berry}, D.~S., ``{JCMT
  Science Archive: Advanced Heterodyne Data Products Pipeline},'' in [{\em
  Astronomical Data Analysis Software and Systems
  XVII}{\nolinebreak\hspace{0.1em}]},  {Argyle}, R.~W., {Bunclark}, P.~S., and
  {Lewis}, J.~R., eds., {\em ASP Conf.\ Ser.} {\bf 394},  565 (2008).

\bibitem{2013ASPC..475..341C}
{Currie}, M.~J., ``{Automated Removal of Bad-Baseline Spectra from ACSIS/HARP
  Heterodyne Time Series},'' in [{\em Astronomical Data Analysis Software and
  Systems XXII}{\nolinebreak\hspace{0.1em}]},  {Friedel}, D.~N., ed., {\em ASP
  Conf. Ser.} {\bf 475},  341 (2013).

\bibitem{JennessACSISDR}
Jenness, T., Currie, M.~J., Cavanagh, B., Berry, D.~S., Leech, J., and Rizzi,
  L., ``{Automated reduction of single-dish heterodyne data from the James
  Clerk Maxwell Telescope using ORAC-DR},'' {\em Astronomy \& Computing}~{\bf
  in preparation} (2014).

\bibitem{2013ascl.soft10001J}
{Jenness}, T., {Economou}, F., {Cavanagh}, B., {Currie}, M., and {Gibb}, A.,
  ``{ORAC-DR: Astronomy data reduction pipeline},'' (2013).
\newblock Astrophysics Source Code Library \ascl{1310.001}.

\bibitem{2003SPIE.4855..322G}
{Graf}, U.~U., {Heyminck}, S., {Michael}, E.~A., et~al., ``{SMART: The KOSMA
  Sub-Millimeter Array Receiver for Two frequencies},'' in [{\em Millimeter and
  Submillimeter Detectors for Astronomy}{\nolinebreak\hspace{0.1em}]},
  {Phillips}, T.~G. and {Zmuidzinas}, J., eds., {\em Proc.\ SPIE} {\bf 4855},
  322--329 (2003).

\bibitem{2005IAUS..235P.275K}
{Kawamura}, A., {Mizuno}, N., {Yonekura}, Y., {Onishi}, T., {Mizuno}, A., and
  {Fukui}, Y., ``{NANTEN2: A Submillimeter Telescope for Large Scale Surveys at
  Atacama},'' in [{\em Astrochemistry: Recent Successes and Current
  Challenges}{\nolinebreak\hspace{0.1em}]},  {\em IAU Symposium} {\bf 235},
  275P (2005).

\bibitem{2008stt..conf..488G}
{Graf}, U.~U., {Honingh}, C.~E., {Jacobs}, K., {Justen}, M., {P{\"u}tz}, P.,
  {Schultz}, M., {Wulff}, S., and {Stutzki}, J., ``{Upgrade of the SMART Focal
  Plane Array Receiver for NANTEN2},'' in [{\em Nineteenth International
  Symposium on Space Terahertz Technology}{\nolinebreak\hspace{0.1em}]},
  {Wild}, W., ed.,  488 (2008).

\bibitem{2013ascl.soft05010G}
{Gildas Team}, ``{GILDAS: Grenoble Image and Line Data Analysis Software},''
  (2013).
\newblock Astrophysics Source Code Library \ascl{1305.010}.

\bibitem{2012SPIE.8452E..04K}
{Kloosterman}, J., {Cottam}, T., {Swift}, B., et~al., ``{First observations
  with SuperCam and future plans},'' in [{\em Millimeter, Submillimeter, and
  Far-Infrared Detectors and Instrumentation for Astronomy
  VI}{\nolinebreak\hspace{0.1em}]},  Holland, W.~S., ed., {\em Proc.\ SPIE}
  {\bf 8452},  845204 (2012).

\bibitem{2010SPIE.7740E..38A}
{Axelrod}, T., {Kantor}, J., {Lupton}, R.~H., and {Pierfederici}, F., ``{An
  open source application framework for astronomical imaging pipelines},'' in
  [{\em Software and Cyberinfrastructure for
  Astronomy}{\nolinebreak\hspace{0.1em}]},  {\em Proc.\ SPIE} {\bf 7740},
  774015 (2010).

\bibitem{2011Kniberg}
Kniberg, H.,  [{\em Lean from the Trenches: Managing Large-Scale Projects with
  Kanban}{\nolinebreak\hspace{0.1em}]}, The Pragmatic Programmers, 1st~ed.
  (2011).
\newblock ISBN-13: 978-1-934356-85-2.

\end{thebibliography}
\end{document}